\documentclass[aps,prb,showpacs,amsmath,amssymb,twocolumn,superscriptaddress]{revtex4-1}
\pdfoutput=1

\usepackage{color}
\usepackage{amsmath}
\usepackage{graphicx}
\usepackage{multirow}

\begin{document}

\newcommand{\bn}{{\bf n}}
\newcommand{\bp}{{\bf p}}   
\newcommand{\br}{{\bf r}}
\newcommand{\bk}{{\bf k}}
\newcommand{\bv}{{\bf v}}
\newcommand{\brho}{{\bm{\rho}}}
\newcommand{\bj}{{\bf j}}
\newcommand{\wk}{\omega_{\bf k}}
\newcommand{\nk}{n_{\bf k}}
\newcommand{\eps}{\varepsilon}
\newcommand{\la}{\langle}
\newcommand{\ra}{\rangle}
\newcommand{\be}{\begin{eqnarray}}
\newcommand{\ee}{\end{eqnarray}}
\newcommand{\intl}{\int\limits_{-\infty}^{\infty}}
\newcommand{\dE}{\delta{\cal E}^{ext}}
\newcommand{\SE}{S_{\cal E}^{ext}}
\newcommand{\dsp}{\displaystyle}
\newcommand{\phit}{\varphi_{\tau}}
\newcommand{\p}{\varphi}
\newcommand{\dphi}{\delta\varphi}
\newcommand{\dbj}{\delta{\bf j}}
\newcommand{\dI}{\delta I}
\newcommand{\dph}{\delta\varphi}
\newcommand{\ua}{\uparrow}
\newcommand{\da}{\downarrow}
\newcommand{\sv}{\sigma_{\alpha}v_F}
\newcommand{\cotV}{\coth\!\left(\frac{e V}{2 T}\right)}

\title{ Noise in the helical  edge channel  anisotropically coupled to a local spin}

\author{K.~E.~Nagaev}
\affiliation{Kotelnikov Institute of Radioengineering and Electronics, Mokhovaya 11-7, Moscow 125009, Russia}

\author{S.\,V. Remizov}
\affiliation{Kotelnikov Institute of Radioengineering and Electronics, Mokhovaya 11-7, Moscow 125009, Russia}
\affiliation{Dukhov Research Institute of Automatics (VNIIA), Moscow 127055, Russia}

\author{D.\,S. Shapiro}
\affiliation{Kotelnikov Institute of Radioengineering and Electronics, Mokhovaya 11-7, Moscow 125009, Russia}
\affiliation{Dukhov Research Institute of Automatics (VNIIA), Moscow 127055, Russia}

\date{\today}

\begin{abstract}
We calculate the frequency-dependent shot noise in the edge states of a two-dimensional topological insulator coupled to 
a magnetic impurity with spin $S=1/2$ of arbitrary anisotropy. If the anisotropy is absent, the noise is purely thermal 
at low frequencies, 
but tends to the Poissonian noise of the full current $I$ at high frequencies. If the interaction only flips the impurity spin 
but conserves those of electrons, the noise at high voltages $eV\gg T$ is frequency-independent. Both the noise and the backscattering 
current $I_{bs}$ saturate at  voltage-independent values. Finally, if the Hamiltonian 
contains all types of non-spin-conserving scattering, the noise at high voltages becomes frequency-dependent again. At low 
frequencies, its ratio to $2eI_{bs}$ is larger than 1 and may reach 2 in the limit $I_{bs}\to 0$. At high frequencies it tends
to 1. 
\end{abstract}

\maketitle

{\bf 1.\;Introduction.}\; 
\label{1}
The suppression of quantized conductance  in the edge states of two-dimensional topological 
insulators remains an outstanding problem. From theoretical considerations, it follows that the electron-spin 
projection in these states is locked to the direction of its momentum and therefore the electrons cannot be backscattered 
if time-reversal symmetry takes place. However experiments revealed that the actual conductance of these states is much smaller
than the theoretical value $e^2/h$ \cite{Hasan10}, which implies a presence of spin-flip processes. Different mechanisms of such processes 
were  proposed, including capture of electrons from the helical edge states into conducting puddles in the bulk of the material 
\cite{Vayrynen13,Essert15},  different combinations of spin-orbit coupling and electron-electron scattering 
\cite{Crepin12,Schmidt12}, or spin-flip scattering at localized 
magnetic moments \cite{Altshuler13,Kurilovich17}. But so far, none of these mechanisms has obtained a definite experimental confirmation.

An efficient tool for determining the mechanism of conduction are measurements of nonequilibrium noise. 
A quantitative measure of this noise is the Fano factor $F = S_I/2eI$, where $S_I$ is the zero frequency noise 
and $I$ is dc current. For noninteracting electrons, $F$ is always smaller 
than 1, so $F>1$ is a signature of interaction effects in the transport. 
So far, most theoretical papers dealing with noise in 2D topological insulators addressed the electron tunneling between the 
helical states of the same chirality at opposite edges of the insulator \cite{Schmidt11,Rizzo13,Edge15,Dolcini15}. The noise 
in the edge states themselves due to
the hyperfine interaction of the electrons with nuclear spins in a presence of nonuniform spin-orbit coupling was 
calculated in Ref. \cite{delMaestro13}. The shot noise that results from the exchange of electrons between the edge states and 
conducting puddles in the bulk of the insulator was calculated in Ref. \cite{Aseev16}. Very recently, V\"ayrynen and Glazman 
considered the current noise  generated by a local magnetic moment coupled to the edge states \cite{Vayrynen17}. These authors 
calculated the noise spectrum for the case of isotropic coupling by extrapolating the Nyquist relation to finite voltages. They 
also presented an expression for the noise in the case of vanishingly small anisotropic coupling at zero frequency and 
high bias.

The extrapolation of the Nyquist relation to finite voltages appears to give the correct results in the case of isotropic 
coupling with impurity, but it does not hold for the anisotropic case.
In this paper, we microscopically calculate the non-equilibrium electrical noise for an arbitrary anisotropy of exchange coupling of the 
edge states to a local spin 1/2. The stochastic equation approach allows us to derive the spectral density in the classical frequencies domain, i.e. for $\hbar\omega\ll T$, and arbitrary voltage biases (the Boltzmann constant is set to 1 throughout 
the paper). Our results coincide with Ref. \cite{Vayrynen17} 
in the limiting cases.

{\bf 2.\;Average current.}\; 
\label{2}
Consider a pair of helical edge states with linear dispersion $\eps(k)
= \pm v_0 k$, where the upper and lower signs correspond to the two possible  $z$ components of electron spin ($\ua$ and $\da$). 
The edge states connect two electron reservoirs kept at constant voltages $\pm V/2$ and are coupled to a magnetic 
impurity via a Hamiltonian \cite{Kimme16}
\begin{multline}
 H_{int} = J_z S_z s_z + J_0\,(S_{+}s_{-} + S_{-}s_{+}) 
 \\+ 
 J_a\,(S_{+} + S_{-})\,s_z 
 \\+
 J_1 S_z\,(s_{+} + s_{-}) + J_2\,(S_{+} s_{+} + S_{-} s_{-}),
 \label{H_int}
\end{multline}
where $S_z$, $S_{\pm}=S_x \pm iS_y$ and $s_z,s_{\pm}$ are the operators of the impurity spin and 
of the spin density of electrons at its location. The first term in Eq. \eqref{H_int} leads only to dephasing of impurity spin.
The second term leads to spin-conserving transitions, the third and fourth change the total $z$ projection of the 
system spin by 1/2, and the last term changes it by 1.

We assume that the coupling is weak \cite{weak} and that there is a sufficiently strong dephasing of the impurity due to 
the first term 
in Eq. \eqref{H_int}. Therefore the dynamics of spin may be described by a master equation for the occupation numbers of the 
spin-up and spin-down states $N_{\ua}$ and $N_{\da}$ of the form
\be\begin{aligned}
 \frac{dN_{\ua}}{dt} = (\Gamma_0^{+} + \Gamma_a + \Gamma_2^{+})\,N_{\da}
  &- (\Gamma_0^{-} + \Gamma_a + \Gamma_2^{-})\,N_{\ua},
  \\
  N_{\da} = 1 &- N_{\ua},
 \label{dN/dt}
\end{aligned}\ee
where the transition rates corresponding to the different terms of Hamiltonian Eq. \eqref{H_int} are given by the Fermi
golden rule. For example, the rates of spin-conserving transitions that result from the second term in Eq. \eqref{H_int}
are
\be
 \Gamma_0^{\pm} = \frac{\alpha_0}{2\pi\hbar} \int d\eps\,f_{\ua,\da}(\eps)\,[1 - f_{\da,\ua}(\eps)],
 \label{Gamma_0} 
\ee
where $\alpha_0 = |J_0|^2/4v_0^2$ and $f_{\ua,\da}(\eps)$ are the distribution functions of spin-up and spin-down
electrons incident on the impurity. The rate of transitions that conserve the projection of electron spins but flip 
the spin of impurity is
\be%gin{multline}
 \Gamma_a = \frac{\alpha_a}{2\pi\hbar} \sum_{\nu=\ua,\da}
 \int d\eps\, f_{\nu}(\eps)\,[1-f_{\nu}(\eps)],
 %\bigl\{ f_{\ua}(\eps)\,[1 - f_{\ua}(\eps)] 
 %\\+ f_{\da}(\eps)\,[1 - f_{\da}(\eps)] \bigr\},
 \label{Gamma_a}
\ee%nd{multline}
where $\alpha_a = |J_a|^2/4v_0^2$. The rates of transitions that flip both the electron and impurity spins in the same
direction $\Gamma_2^{\pm}$ coincide with $\Gamma_0^{\mp}$ up to the replacement of $\alpha_0$ by $\alpha_2 = |J_2|^2/4v_0^2$.

The electrical current through the edge states is given by the expression
\begin{multline}
 I = I_{\ua}^{in} + I_{\da}^{in} - e\,(\Gamma_0^{+}\,N_{\da} - \Gamma_0^{-}\,N_{\ua})
 \\
 + e\,(\Gamma_1^{+} - \Gamma_1^{-}) 
 + e\,(\Gamma_2^{+}\,N_{\da} - \Gamma_2^{-}\,N_{\ua}),
 \label{I-1}
\end{multline}
where the currents injected into the edge states from the left and right reservoirs are given by equations
\be
 I_{\ua,\da}^{in} =  \pm\frac{e}{2\pi\hbar} \int d\eps\,f_{\ua,\da}(\eps)
 \label{I_in-1}
\ee
and the rest of terms describe spin-flip backscattering of electrons from the impurity, which is either accompanied by its
spin flip or not. The rates of scattering events that do not change the impurity spin  $\Gamma_1^{\pm}$ are obtained from
$\Gamma_0^{\mp}$ by replacing $\alpha_0$ by $\alpha_1 = |J_1|^2/4v_0^2$.

The dc current is easily obtained from the stationary solution of Eqs. \eqref{dN/dt} and \eqref{I-1}. 
Because of weak coupling, one may calculate the transition rates using the unperturbed distribution 
functions of electrons $f_{\ua}(\eps) = f_0(\eps-eV/2)$ and $f_{\da}(\eps) = f_0(\eps+eV/2)$, where $f_0(\eps)$
is the Fermi distribution.
In the most general case, it is of the form
\begin{multline}
 I = \frac{e^2 V}{4\pi\hbar}\,
 (2 - \alpha_0  - 2\alpha_1 -  \alpha_2)
\\ 
 + 
 \frac{e^2 V}{4\pi\hbar}\,
 \frac{(\alpha_0-\alpha_2)^2\,eV \cotV}
      {4\alpha_a T + (\alpha_0 + \alpha_2)\,eV \cotV}.
 \label{I-2} 
\end{multline}
The shape of negative correction to the current in the absence of scattering $I_{bs}=e^2V/2\pi\hbar -I$ strongly depends on 
the coupling-parameter anisotropy in the Hamiltonian Eq. \eqref{H_int}. If $\alpha_a=\alpha_1=\alpha_2=0$, $I_{bs}$ is also
zero for any $V$. If $\alpha_a \ne 0$ but $\alpha_1=\alpha_2=0$, $I_{bs}$ is linearly proportional to the voltage at 
$eV \ll T$ but saturates to a finite value in the opposite limit. Finally if either $\alpha_1$ or $\alpha_2$ is nonzero, the 
backscattering current scales linearly with $V$ even at high voltages. 

{\bf 3.\;Equations for the fluctuations.}\; 
\label{3}
Because of weak coupling between the electrons and the impurity, the dynamics of
the system may  be treated as a Markov random process and described by differential stochastic equations. Basically, there are
two sources of noise in this system. One of them is the random injection of electrons in the edge states from the reservoirs,
i.~e. the occupation-number noise. Taken alone, this random injection would result in the Nyquist noise of a perfect quantum
channel with a spectral density $2e^2T/\pi\hbar$. Out of equilibrium, there is also another source of noise related with the 
randomness of electron-impurity spin-flip scattering, i.~e. the partition noise \cite{Blanter00}. The interference between these two sources 
of noise shapes its temperature and voltage dependence.
The stochastic equation for the impurity spin-up occupation number is of the form
\begin{multline}
 \frac{d\delta N_{\ua}}{dt} =
 -(\Gamma_0^{+} + \Gamma_0^{-} + 2\Gamma_a + \Gamma_2^{+} + \Gamma_2^{-})\,\delta N_{\ua}
\\
 + N_{\da}\,(\delta\Gamma_0^{+} + \delta\Gamma_2^{+})
 - N_{\ua}\,(\delta\Gamma_0^{-} + \delta\Gamma_2^{-})
\\
 + (N_{\da} - N_{\ua})\,\delta\Gamma_a + \xi_0 + \xi_a + \xi_2,
\label{L-spin}
\end{multline}
and the corresponding Langevin equation for the fluctuation of the current is 
\begin{multline}
 \delta I = \delta I_{\ua}^{in} + \delta I_{\da}^{in}
 + e\,(\Gamma_0^{+} + \Gamma_0^{-} - \Gamma_2^{+} - \Gamma_2^{-})\,\delta N_{\ua}
\\
 - eN_{\da}\,(\delta\Gamma_0^{+} - \delta\Gamma_2^{+})
 + eN_{\ua}\,(\delta\Gamma_0^{-} - \delta\Gamma_2^{-})
\\
 + e\,       (\delta\Gamma_1^{+} - \delta\Gamma_1^{-})
 - e\,(\xi_0 - \xi_1 - \xi_2),
\label{L-current}
\end{multline}
where the symbol $\delta\ldots$ denotes the fluctuation of the corresponding quantity and 
the external sources $\xi_0$, $\xi_a$, $\xi_1$, and $\xi_2$ are the external sources, which are delta-correlated
in time and related with the scattering processes described by $\Gamma_0^{\pm}$, $\Gamma_a$, $\Gamma_1^{\pm}$, and 
$\Gamma_2^{\pm}$, respectively. Note that $\xi_0$ enters into Eqs. \eqref{L-spin} and \eqref{L-current} with opposite 
signs because this type of scattering decreases the number of right-moving electrons as the impurity spin turns upward.
In contrast to this, $\xi_2$ enters into both equations with the same sign because this type of scattering simultaneously
increases the number of right-movers and the projection of the impurity spin on the $z$ axis.
In the limit of weak electron-impurity coupling, the spectral density of these sources is twice the sum of scattering
fluxes in both directions \cite{Nagaev15}, i.~e.
\be\begin{aligned}
 S_{0} = 2\,(\Gamma_0^{+}\,N_{\da} + \Gamma_0^{-}\,N_{\ua}),
 \quad
 S_{a} = 2\,\Gamma_a, \qquad{}
 \\
 S_{1} = 2\,(\Gamma_1^{+} + \Gamma_1^{-}),
 \quad
 S_{2} = 2\,(\Gamma_2^{+}\,N_{\da} + \Gamma_2^{-}\,N_{\ua}).
\end{aligned}\label{S_xi}\ee
The extraneous sources related with different scattering processes are uncorrelated.

While the partition noise is described by the extraneous sources in Eqs. \eqref{L-spin} and \eqref{L-current}, 
the occupation-number noise comes into play through the fluctuations of the distribution functions of injected electrons
$\delta f_{\ua,\da}(\eps)$, which result in the fluctuations of the injected currents \eqref{I_in-1} and the
scattering rates, e.~g.
\be
 \delta\Gamma_0^{+} = \frac{\alpha_0}{2\pi\hbar} \int d\eps\, [(1-f_{\da})\,\delta f_{\ua}  - f_{\ua}\,\delta f_{\da}].
 \label{dG}
\ee
The correlation function of $\delta f_{\ua,\da}(\eps)$ is well known and equals \cite{deJong96}
\begin{multline}
 \la \delta f_{\sigma}(t,\eps)\,\delta f_{\sigma'}(t',\eps')\ra = 2\pi\hbar\,\delta_{\sigma\sigma'}\,
 \delta(t-t')\,\delta(\eps-\eps')\,
 \\ \times f_{\sigma}^{in}(\eps)\,[1 - f_{\sigma}^{in}(\eps)],
 \label{df-df}
\end{multline}
where $\sigma= (\ua,\da)$. These fluctuations are uncorrelated with the extraneous sources in Eqs. \eqref{L-spin} and 
\eqref{L-current} because they originate from the depth of reservoirs due to the finite temperature in them. 
Equation \eqref{L-spin} has to be solved for $\delta N_{\ua}$ and the result has to be substituted into 
Eq.~\eqref{L-current} together with $\delta N_{\da} = -\delta N_{\da}$. Multiplying the Fourier transform of 
Eq. \eqref{L-current} by its complex conjugate and making use of Eqs. \eqref{S_xi} and \eqref{df-df}, one obtains 
the spectral density
of current noise. Because the coupling is assumed to be weak, one has to keep only the terms up to linear order in 
constants $\alpha$.
The general equation for the nonequilibrium noise is too cumbersome to be presented here, and 
therefore we give it only for some particular cases.

{\bf 4.\;The equilibrium response.}\; 
\label{4}
First of all, we calculate the equilibrium linear response of the edge states 
coupled to an impurity to a voltage oscillating with frequency $\omega$. It is easily obtained from Eqs. \eqref{L-spin} 
and \eqref{L-current} by setting
$\delta f_{\ua,\da}(\eps,t) = \mp(eV/2)\exp(-i\omega t)\,df_0/d\eps$ and dropping the extraneous sources in them. In
this case, the time-averaged spin-up and spin-down occupancies of the impurity are equal to 1/2, and the frequency-dependent
complex conductance of the system is given by
\begin{multline}
 G(\omega) = \frac{e^2}{2\pi\hbar}
 \biggl[ 1 - \alpha_1
  -\frac{\alpha_0}{2}\,
%\\ \times
   \frac{ -2\pi i\,\hbar\omega + 4\,(\alpha_a + \alpha_2)\,T }{ -2\pi i\,\hbar\omega + 2\,%(\alpha_0 + 2\alpha_a + \alpha_2)\,T }
   \alpha_{\Sigma}\,T}
\\
  -\frac{\alpha_2}{2}\,
   \frac{ -2\pi i\,\hbar\omega + 4\,(\alpha_a + \alpha_0)\,T }{ -2\pi i\,\hbar\omega + 2\,%(\alpha_0 + 2\alpha_a + \alpha_2)\,T }
   \alpha_{\Sigma}\,T}
 \biggr],
 \label{resp_eq}
\end{multline}
where $\alpha_{\Sigma} = \alpha_0 + 2\alpha_a + \alpha_2$. Unless $\alpha_0=\alpha_2$, the real part of $G$ monotonically 
decreases with $\omega$ to a 
finite value, while its imaginary part tends to  zero both at $\omega=0$ and high frequencies with a maximum at
$\omega \sim \alpha_{\Sigma}T/\hbar$. It may be verified that the equilibrium noise calculated by our method 
satisfies the Nyquist theorem $S_I(\omega) = 4T\,{\rm Re}\, G(\omega)$ for arbitrary coupling constants.

{\bf 5.\;Rotationally symmetric coupling.}\; 
\label{5}
If the electron-impurity coupling is rotationally symmetric with respect to the $z$ axis, the only 
nonzero coupling constant is $J_0$ and the only nonzero
scattering rate is $\Gamma_0^{\pm}$. In this case, the total spin of the electrons and the impurity is conserved
by the scattering, and therefore the dc current is not affected by it because the impurity cannot increase the $z$
projection of its spin to infinity. For this reason, the spectral density of noise at zero frequency is not affected
by the scattering either. At nonzero frequencies, it is of the form
\be\begin{aligned}
 S_I(\omega) = \frac{e^2}{\pi\hbar}
 \left[2\,T - \frac{\alpha_0\,\omega^2}{\omega^2 + \tau_0^{-2}}\,\frac{eV}{\sinh(eV/T)} \right],
 \\
 \tau_0^{-1} = \alpha_0\,\frac{eV}{2\pi\hbar}\,\coth\!\left(\frac{eV}{2T}\right).\qquad\quad{}
 \label{S_I-rs}
\end{aligned}\ee
This is essentially the same result that was obtained in Ref. \cite{Vayrynen17} by extrapolating the fluctuation-dissipation
relation into the nonequilibrium region. The reason why this extrapolation is valid in this particular case is that
for an isotropic coupling, the effect of finite voltage difference between the reservoirs is equivalent
to an external magnetic field applied to the impurity along the $z$ axis \cite{Tanaka11}. Hence the nonequilibrium system may be
mapped onto an equilibrium one and the fluctuation-dissipation relation may be used.

{\bf 6.\;Point-like impurity.}\; 
\label{6}
The coupling may be anisotropic even for a point-like impurity and isotropic exchange interaction with bulk electrons 
if a finite spatial width of the edge state is taken into account and the impurity is located away from its center 
in the $z$ direction \cite{Kimme16}. In this case, the coupling constant $J_a$ { is nonzero
as well as $J_0$,  so both  $\Gamma_a$ and $\Gamma_0^{\pm}$ have to be taken into account.}   The scattering process 
described by $\Gamma_a$ leads to a relaxation of the impurity spin similarly to its 
contact with an external spin bath, and therefore the scattering correction to the dc current is nonzero. However 
$\Gamma_a$ is proportional to $T/\hbar$, while $\Gamma_0^{+}$ is proportional to $eV/\hbar$ at $eV\gg T$. This is why 
the backscattering current initially grows with voltage but eventually saturates at
$I_{bs} = \alpha_aeT/\pi\hbar$. The nonequilibrium noise shows a similar behavior.
At a given voltage, it monotonically decreases with increasing frequency to a finite value. A typical voltage dependence
of the zero-frequency and high-frequency noise is shown in Fig. 1. In the zero-voltage limit, the noise decreases from
\be
 S_I(0) = \dfrac{2e^2}{\pi\hbar} \left(1 - \dfrac{\alpha_a\alpha_0}{\alpha_0 + 2\alpha_a}\right) T
\label{S_I-a-zf}
\ee
\begin{figure}[t]
\center
\includegraphics[width=0.8\linewidth]{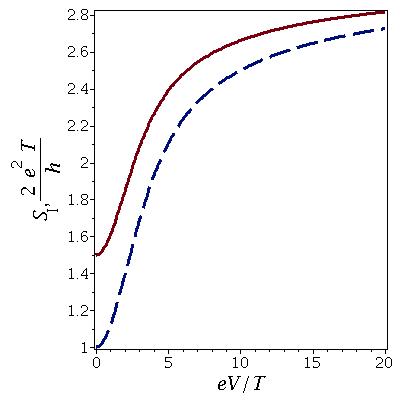}
\caption{Fig.~\ref{figure:fig1}.The voltage dependence of $S_I(0)$ (solid curve) and $S_I(\infty)$ (dashed curve) for a 
coupling with a point-like impurity with $\alpha_0=1$ and $\alpha_a=0.5$.}
\label{figure:fig1}
\end{figure}
at  low frequency to 
\be
 S_I(\infty) = \dfrac{e^2}{\pi\hbar}\,(2 - \alpha_0)\,T
\label{S_I-a-hf}
\ee
at high frequencies. At high voltages the noise is frequency-independent and equals to
\be
 S_I = \dfrac{2e^2}{\pi\hbar}\,(1 + \alpha_a)\,T.
\label{S_I-a-hV}
\ee
Note the change of sign of the inelastic correction to the noise with respect to Eqs. \eqref{S_I-a-zf}  and
\eqref{S_I-a-hf}. The Fano factor of the excess noise with respect to the backscattering current
$F_{bs} \equiv (S_I - 2e^2T/\pi\hbar)/2eI_{bs}$ is unity. This suggests that the backscattering of different electrons 
from the impurity is totally uncorrelated.

{\bf 7.\;Coupling of arbitrary symmetry.}\; 
\label{7}
If the impurity has a finite size, all the coupling parameters in the Hamiltonian \eqref{H_int} may be nonzero. 
The scattering rate $\Gamma_1$ leads to a backscattering of electrons without flipping the impurity spin,
while the rates $\Gamma_0^{\pm}$ and $\Gamma_2^{\pm}$ describe electron backscattering accompanied by flips of the impurity 
spin in the opposite directions, which partly compensate each other. As the three scattering rates are proportional
to $eV/\hbar$ at $eV\gg T$, the backscattering current and the current noise also increase proportionally to the
voltage. At low voltages, the zero-frequency noise is given by the Nyquist formula
\be
 S_I(0)=
 \dfrac{2e^2}{\pi\hbar} 
 \left[
   1 - \alpha_1 - \dfrac{\alpha_a\,(\alpha_0+\alpha_2) + 2\alpha_0\alpha_2}{\alpha_0 + 2\alpha_a + \alpha_2}
 \right] T.
 \label{S_I-eq-w0}
\ee
As the voltage increases, the scattering correction to the noise becomes positive and assumes the form
\be
 S_I(0) = \dfrac{e^2}{\pi\hbar}
 \left[ 
  \alpha_1 + 4\,\dfrac{\alpha_0\alpha_2\,(\alpha_0^2 + \alpha_2^2)}{(\alpha_0+\alpha_2)^3} 
 \right] eV.
 \label{S_I-shot-w0}
\ee
Together with Eq. \eqref{I-2} it results in the Fano factor with respect to backscattering current
\be
 F_{bs} = 
  \frac{ \alpha_1\,(\alpha_0+\alpha_2)^3 + 4\alpha_0\alpha_2\,(\alpha_0^2 + \alpha_2^2)}
      { [\alpha_1\,(\alpha_0+\alpha_2) + 2\alpha_0\alpha_2](\alpha_0+\alpha_2)^2 }.
 \label{F_bs}
\ee
\begin{figure}[t]
\center
\includegraphics[width=0.8\linewidth]{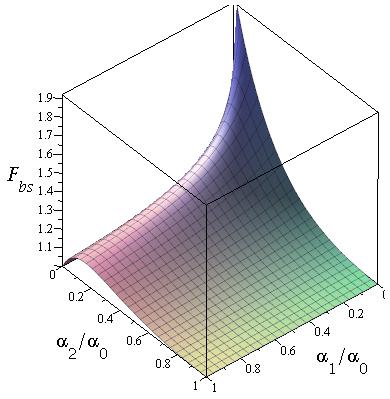}
\caption{Fig.~\ref{figure:fig2}. A 3D plot of the Fano factor $F_{bs}$ as a function of $\alpha_1/\alpha_0$
and $\alpha_2/\alpha_0$.}
\label{figure:fig2}
\end{figure}
As well as the dc current Eq. \eqref{I-2}, this equation is symmetric with respect to $\alpha_0$ and
$\alpha_2$, but there are reasons to believe that both $\alpha_1$ and $\alpha_2$ are smaller than $\alpha_0$ in realistic 
systems \cite{Kimme16}.
A 3D plot of Eq. \eqref{F_bs} as a function of $\alpha_1/\alpha_0$ and $\alpha_2/\alpha_0$ is shown in Fig. 2.
Depending on these ratios, it varies between 1 and 2 and reaches maximum in the limit $\alpha_1 \ll \alpha_2 \ll \alpha_0$.
The increase of $F_{bs}$ above 1 suggests that the events of electron backscattering from the impurity are correlated.
Indeed, if either $\alpha_2$ or $\alpha_0$ would be zero, the impurity would be completely polarized and no electron
backscattering would be possible. If both of them are nonzero, the scattering events with the smaller rate would destroy
this polarization and favor the scattering event with larger rate that will flip the impurity spin in the opposite direction
\cite{Vayrynen17}. Therefore the electron backscattering events  take place in pairs and the zero-frequency noise is doubled. 
Note that $F_{bs}$ decreases to unity if $\alpha_2=\alpha_0$. A similar increase of the Fano factor above unity was observed
in resonant tunneling via interacting localized states \cite{Safonov03}. 

The dispersion of the noise takes place at 
$\omega \sim\Gamma_0^{+} + \Gamma_2^{-}$. At much higher frequencies and at $eV\gg T$, 
\be 
 S_I(\infty) = \dfrac{e^2}{\pi\hbar} 
 \left( \alpha_1 + 2\,\dfrac{\alpha_0\alpha_2}{\alpha_0+\alpha_2} \right) eV
 \label{S_I-shot-hf}
\ee
is always smaller than $S_I(0)$ and its ratio to $2eI_{bs}$ is one. At $\alpha_1 \ll \alpha_2 \ll \alpha_0$, the frequency
dependence of spectral 
density is consistent with a picture of a random sequence of current pulses of the form
\be
 I_p(t) = e\,\bigr[ \sqrt{2}\,\delta(t) + (2-\sqrt{2})\,\Gamma_0\,\Theta(t)\exp(-\Gamma_0 t)\bigr],
 \label{pulse}
\ee
which carry a charge of $2e$ each.

{\bf 8.\;Conclusion.}\; 
\label{8}
So far, there were not many experimental results on electrical noise in the edge states of 2D topological insulators.
The only paper we are aware of is Ref. \cite{Tikhonov15}, which reported the conductance much smaller than $e^2/\hbar$ and the Fano
factor smaller than one. Still our results may be compatible with these findings if one assumes that the backscattering
is caused by many local moments randomly distributed along the edge states. In this case, the Fano factor with respect to
the transport current would reduce to 1/3, as in conventional diffusive conductors. To test the current theory, one could
controllably implant magnetic impurities like Mn near the edges of a topological insulator.

In summary, we have calculated the nonequilibrium electric noise in a pair of edge states in a 2D topological insulator
with coupling to a magnetic impurity of arbitrary anisotropy at classical frequencies. For the rotationally symmetric 
coupling, the noise deviates from the equilibrium one only at finite frequencies. If a point-like impurity is located
away from the middle plane of the 2D insulator, both the backscattering current and nonequlibrium noise increase with
voltage to a finite value proportional to the temperature. The Fano factor with respect to the backscattering
current is unity in this case. If the coupling has an arbitrary symmetry, the noise grows linearly with voltage
and the Fano factor ranges between 1 and 2 depending on the ratio between different coupling parameters. For a particular
choice of these parameters, the backscattering current can be viewed as random sequence of pulses of asymmetric shape, 
each carrying a charge $2e$.

This work was supported by Russian Science Foundation under Grant No. 16-12-10335.


\begin{thebibliography}{99}
\bibitem{weak}
This implies that the Kondo-like effects are ignored, i.~e. that we are well above the Kondo temperature.
\end{thebibliography}
\end{document}